\documentclass[aps,prb,groupedaddress,superscriptaddress,showkeys,showpacs,twocolumn,nolongbibliography]{revtex4-2}
\usepackage{amsfonts}
\usepackage{amsmath}
\usepackage{amssymb}
\usepackage{graphicx}
\usepackage{mathcomp}
\usepackage{stmaryrd}
\usepackage[cal=boondox]{mathalfa}

\usepackage{amsmath}

\usepackage{listings}
\usepackage{color}

\definecolor{dkgreen}{rgb}{0,0.6,0}
\definecolor{gray}{rgb}{0.5,0.5,0.5}
\definecolor{mauve}{rgb}{0.58,0,0.82}

\usepackage{mathtools}
\usepackage{braket}

\usepackage[colorlinks=true,allcolors=blue]{hyperref}
\usepackage{graphicx}

\usepackage{savesym}
\usepackage{bm}
\savesymbol{iint}
\usepackage{wasysym}
\restoresymbol{WSS}{iint}

\newcommand {\R}{\mathbb{R}}

\begin{document}

\title{Application of an upsampling algorithm to quantum state preparation of continuous and discrete probability distributions}

\author{R. P. Erickson}
\email[]{Electronic address: Robert.Erickson2@wellsfargo.com}
\affiliation{Wells Fargo}

\date{\today}

\begin{abstract}
Upsampling of time series data is often associated with classical fast Fourier transform analysis, but it also can be employed in the preparation of a quantum state vector. The quantum circuit of preparation derived from functional data sampled in this way exhibits exponential gate depth, like other divide-and-conquer algorithms, but is transformable to an equivalent circuit of polylogarithmic gate depth. In this study, we derive a polylogarthmic upsampling algorithm for construction of a state vector with amplitudes that are square roots of probabilities sampled from a continuous probability distribution having support over the entire real line. As examples, we prepare state vectors associated with Gaussian and Laplace distributions; state vectors for other continuous probability distributions such as Cauchy and Student's t follow in a similar manner. We also extend the algorithm to include preparation of a state vector whose amplitudes are square roots of an arbitrary distribution of discrete probabilities. Our analyses focus on univariate distributions, but can be extended to multivariate forms. The polylogarithmic upsampling algorithm has financial and scientific application.
\end{abstract}

\maketitle

\section{Introduction\label{sec:introduction}}
Many financial, scientific, and machine-learning quantum computing applications require loading of classical information into a state vector. Repeated use of this entangled quantum state in a calculation necessitates reloading due to the no-cloning restriction imposed by the nature of quantum mechanics \cite{Nielsen_Chuang_2010}. Additionally, limitations exist on both dimension of state vector and time to completion of the underlying calculation due to quantum decoherence, external noise and stray fields, and scaling issues that inhibit control and precision, such as the overhead of large numbers of physical qubits needed to implement quantum error correction schemes. These restrictions impose a high level of efficiency on any quantum algorithm used to prepare the state vector \cite{Aaronson2015,*Biamonte2017}. To address these challenges, recent efforts have included the use of numerical integration \cite{Grover2002,*CarreraVazquez2021}, quantum generative adversarial networks \cite{Zoufal2019,*Zhu2022}, black-box (oracle) approaches \cite{Sanders2019,*Bausch2022}, matrix product states \cite{PhysRevA.101.010301,*Plekhanov2022variationalquantum,*akhalwaya2023modularenginequantummonte,*Iaconis2024,*sano2024}, and other methods emphasizing fault tolerance \cite{Sun2023,*Zhang2022,*Rattew2022,*McArdle2022}.

In this study we apply an upsampling method of data capture, most often associated with classical fast Fourier transform analysis of discrete samples of time series, to the construction of a quantum algorithm that can be used to prepare a state vector whose amplitudes are obtained from the discretization of a continuous function. The algorithm derived is a divide-and-conquer algorithm \cite{Cormen2009} composed of controlled y-rotation gates \cite{Mottonen2005}, with exponential gate depth, but it can be transformed to an equivalent circuit of polylogarthmic gate depth \cite{Araujo2021} using a method of quantum forking \cite{Park2019b}. In our analysis we consider the specific univariate examples of Gauss and Laplace; application to other continuous probability distributions, such as Cauchy and Student's t, follows from the formulas presented. We also extend our study to include the case of distributions of discrete probabilities.

In our method of quantum state preparation, we adapt elements of the approach of Kitaev and Webb \cite{kitaev2009wavefunction}, in particular, defining a periodic equivalent of a continuous probability density function (PDF). The quantum state vector is prepared such that the probability of measurement of a given multi-qubit state is proportional to the value of a point on the periodic function. We capture the functional dependence of the PDF within the computational resource using conventional time-series-like upsampling, defining a sampling interval (and a limiting Nyquist frequency) as a function of the employed number of qubits.

Specifically, if a univariate PDF is supported on the entire real line then it can be denoted by continuous function $P(x)$, where $x\in\R$ and the probability of occurrence of $x$ on the interval of $x$ to $x+dx$ is $P(x)\, dx$. For a computational resource consisting of $n$ qubits, we define unsigned integers $0\le i_n<2^n$ and represent them in terms of a set of bit values $i^{(n)}_m\in\left\{ 0,1 \right\}$, where $0\le m<n $, such that
\begin{equation}	\label{eq:integer}
i_n=\sum_{m=0}^{n-1} 2^m \, i^{(n)}_m \; ; \quad i_0 \equiv 0 .
\end{equation}
In the little endian convention, used throughout this study, an $n$-qubit state can be defined as $\ket{i_n}=\ket{i^{(n)}_{n-1},\dots,i^{(n)}_0}$, where $i^{(n)}_{n-1}$ is the most significant bit. 

Unlike \cite{kitaev2009wavefunction}, we capture $P(x)$ within a sampling window, $w$, and map only integers $i_n$ (supporting the $n$-qubit system) to real values $x_n$ via the linear formula
\begin{equation}	\label{eq:xmap}
x_n=x_o + \Delta x_n \cdot i_n \; ;
\end{equation}
\begin{equation}	\label{eq:xmap-2}
\Delta x_n = \frac{w}{2^n} , \quad 
x_o = \bar{x} + w \left( \zeta - 1 \right) / 2 ,
\end{equation}
where $P(\bar{x})=\max{P(x)}$, $\Delta x_n$ is the sampling interval, and $f_n=1/(2\Delta x_n)$ is the Nyquist frequency. 

\begin{figure}[h]
\includegraphics[width=240pt, height=240pt]{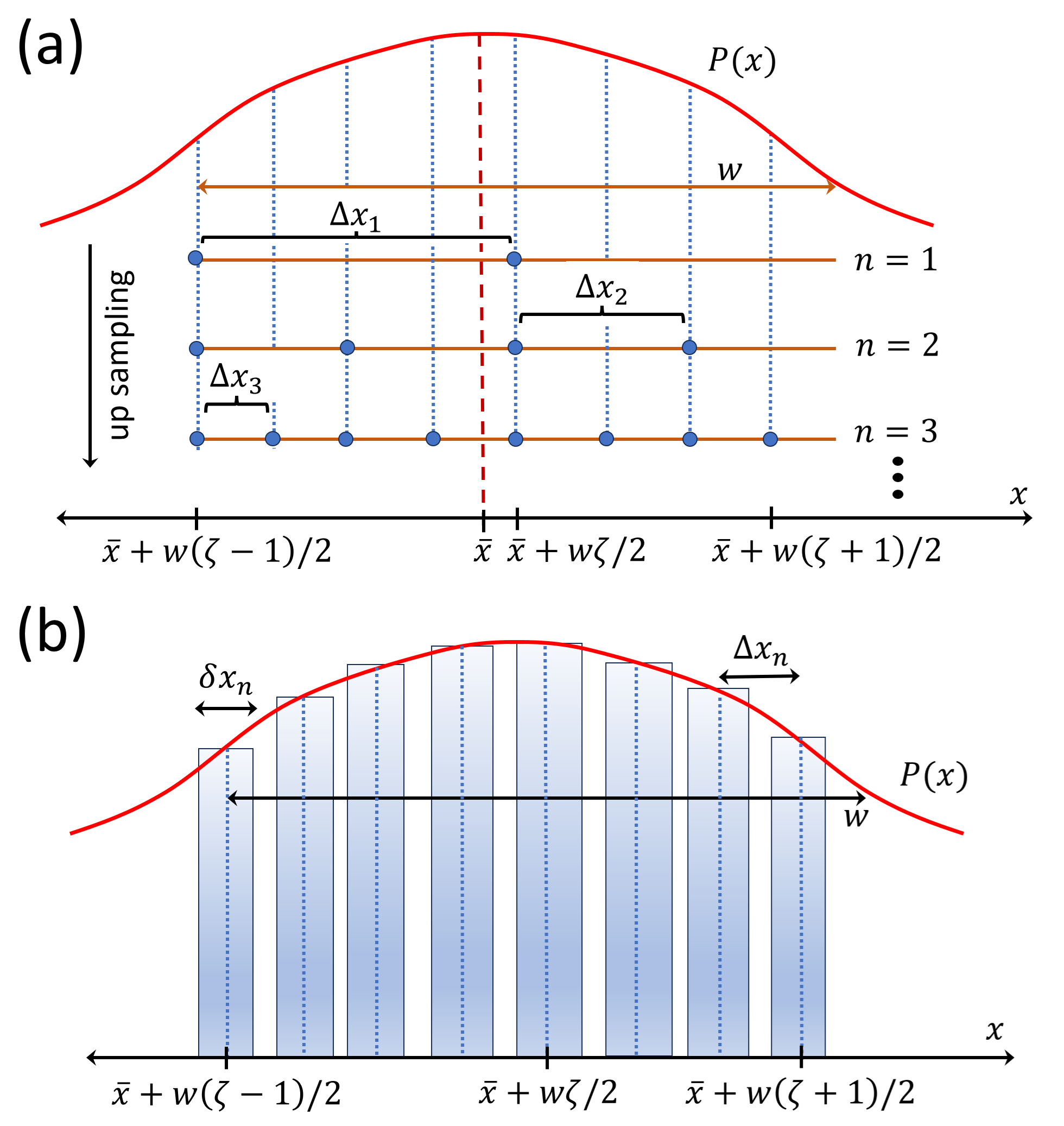}
\caption{\label{fig1} (a) Upsampling capture of PDF, $P(x)$ (solid red curve), as a function of increasing number of qubits, $n$, for fixed values of window width, $w$, and parameter $\zeta$. For each $n$, the sampling window is centered about $x=\bar{x}+w \zeta/2$, where $P(\bar{x})=\max{P(x)}$ and contains $2^n$ points (blue dots) uniformly separated in distance by $\Delta x_n$, the sampling interval. (b) Plot of $P(x)$ vs $x$ showing how sample probabilities, ${ \left| \braket{i_n|\psi_n} \right| }^2$ of (\ref{eq:probability}), corresponding to a system of $n$ qubits, are represented by areas under curve $P(x)$. Note that normalization factor, $\delta x_n$, and sampling interval, $\Delta x_n$, are different quantities.}
\end{figure}

Specific to our approach, the parameter $\zeta$, where $0\le\zeta <1/2^{n-1}$, can be chosen at random and introduces a small shift in a sampling window that would otherwise be centered at $\bar{x}$. Two state vectors of equal dimension prepared with different values of $\zeta$ will consist of distinctly different sets of amplitudes/probabilities, preventing aliasing when resampling the PDF \footnote{If $\zeta$ were varied continuously from 0 to $1/2^{n-1}$ then ergodic coverage of the entire PDF over the breadth of the sample window would be achieved.}. Figure~\ref{fig1}(a) illustrates our upsampling approach as a function of increasing $n$, for fixed values of $w$ and $\zeta$.

We write the state vector to be prepared, $\ket{\psi_n}$, as a superposition of qubit states, $\ket{i_n}$, viz.
\begin{equation}	\label{eq:psi}
\ket{\psi_n} = \sum_{i_n} \xi_n(i_n) \, \ket{i_n} ,
\end{equation}
where $\xi_n(i_n)$ is an amplitude, and the sum is over all unsigned integers $0\le i_n<2^n$. Similar to \cite{kitaev2009wavefunction}, invoking the map of (\ref{eq:xmap}) so that $\xi_n(i_n)=\xi_n(x_n(i_n))=\xi_n(x_n)$, we define, for any $x\in\R$, that
\begin{equation}	\label{eq:xi}
\xi_n(x) = \sqrt{ \delta x_n \sum_{j=-\infty}^\infty P(x+j w) } ,
\end{equation}
such that $\xi_n(x)$ is periodic in $w$, viz.
\begin{equation}	\label{eq:xi-2}
\xi_n(x+w) = \xi_n(x) .
\end{equation}
Then, requiring $\braket{\psi_n|\psi_n}=1$, the normalization factor, $\delta x_n$, satisfies the constraint
\begin{equation}	\label{eq:dxconstraint}
\sum_{j=-\infty}^\infty \sum_{x_n} P(x_n+j w) \, \delta x_n = 1 .
\end{equation}

Several observations can be made about the above construction of $\ket{\psi_n}$. As stated in (\ref{eq:xi-2}), the first observation is that $\xi_n(x_n)$ of (\ref{eq:xi}) is periodic in $x_n$ with period $w$. A second observation is the limiting behavior of (\ref{eq:dxconstraint}), viz.
\begin{multline}
\lim_{n\rightarrow\infty}\sum_{j=-\infty}^\infty \sum_{x_n} P(x_n+j w) \, \delta x_n \\
= \sum_{j=-\infty}^\infty \int_{-w/2}^{w/2} P(x+j w) \, dx = \int_{-\infty}^{\infty} P(x) \, dx = 1 ,
\end{multline}
where $\delta x_n$ becomes infinitesimally small as $n\rightarrow\infty$, such that the sum over $x_n$ can be replaced by an integral. A third observation, due to the support of $P(x)$ over all $x\in\R$, is that when $w$ is chosen sufficiently large then
\begin{equation}	\label{eq:pdf-condition}
\sum_{j=-\infty}^\infty P(x_n+j w) \cong P(x_n) ,
\end{equation}
to very good approximation, such that
\begin{equation}	\label{eq:approximation}
\xi_n(x_n) \cong \sqrt{ P(x_n) \, \delta x_n } ,
\end{equation} 
\begin{equation}	\label{eq:probability}
{ \left| \braket{i_n|\psi_n} \right| }^2 \cong P(x_n) \, \delta x_n .
\end{equation}
A fourth observation is that the quantities $\delta x_n$ and $\Delta x_n$ are not the same; the former is a normalization factor while the latter is the sampling interval. Figure~\ref{fig1}(b) provides a graphical illustration of these last two observations.

In what follows we present the upsampling algorithm for preparing $\ket{\psi_n}$ of (\ref{eq:psi}), applicable to the Gaussian and Laplace probability distributions, specifically. We then extend our approach to show how a state vector corresponding to a set of discrete probabilities can be prepared by this same algorithm. Next, we apply the concept of quantum forking \cite{Park2019b} to our algorithm to transform it to a polylogarithmic form, following on with several example calculations. We conclude with several final remarks.

\section{Theory\label{sec:theory}}
We first derive the upsampling quantum algorithm applicable to continuous probability distributions. We then use this result to show how state vectors associated with discrete probabilities can be prepared. Lastly, we provide a brief overview of the quantum forking algorithm as it applies to upsampling.

\subsection{The Upsampling Quantum Circuit\label{subsec:upsampling}}
In the first step of the derivation, we write the normalization factor, $\delta x_n$, in its simplest form using the summation identity
\begin{multline}	\label{eq:sumIdentity}
\sum_{j=-\infty}^\infty P( x + j y ) = \sum_{j=-\infty}^\infty P\left( x + 2j y \right) \\
+ \sum_{j=-\infty}^\infty P\left( x + \left( 2j + 1 \right) y \right) \; ; \quad x,y\in\R .
\end{multline}
Solving for $\delta x_n$ in (\ref{eq:dxconstraint}) we initially have 
\begin{equation}	\label{eq:deltax-1}
\delta x_n = \frac{1}{ \sum_{j=-\infty}^\infty \sum_{i_n} P(x_n(i_n)+j w) } ,
\end{equation}
but the sum over $i_n$ can be broken up into pairs of terms involving $P(x_n+j w)$ and $P(x_n+\frac{1}{2}w+j w)$. In other words, dividing the sum into two parts with respect to the most significant bit, we can write
\begin{multline}	\label{eq:deltax-2}
\delta x_n = \Bigg\{ \sum_{j=-\infty}^\infty \sum_{i_{n-1}} \Big[ P(x_n(i_{n-1})+j w) \\
+ P(x_n(i_{n-1})+\left( 2j + 1 \right) w/2) \Big] \Bigg\} ^{-1} ,
\end{multline}
then invoke (\ref{eq:sumIdentity}) to obtain the reduction
\begin{equation}	\label{eq:deltax-3}
\delta x_n = \frac{1}{ \sum_{j=-\infty}^\infty \sum_{i_{n-1}} P(x_n(i_{n-1})+j w/2) } .
\end{equation}
We repeat this process, breaking up the sum over $i_{n-1}$ into terms involving $P(x_n+j w/2)$ and $P(x_n+\frac{1}{4}w+j w/2)$, and again applying (\ref{eq:sumIdentity}). Performing this action $n-1$ times, we arrive at the final expression
\begin{equation}	\label{eq:deltax-1}
\delta x_n = \frac{1}{ \sum_{j=-\infty}^\infty P(x_o+j w/2^n) } .
\end{equation}
\begin{figure*}
\includegraphics[width=500pt, height=120pt]{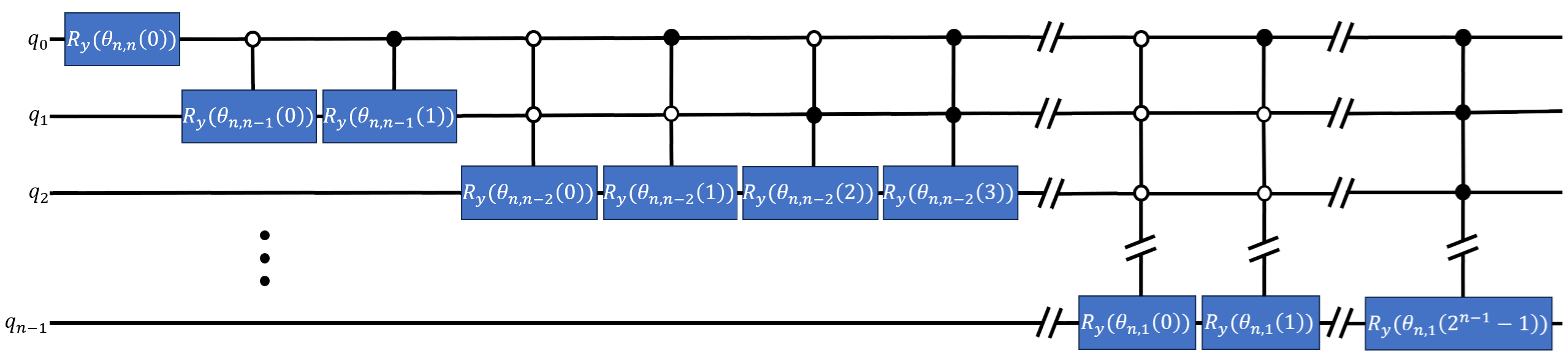}
\caption{\label{fig2} Circuit diagram for the $n$-qubit upsampling algorithm, parameterized by rotation angles $\theta_{n,m}(i)$, as defined in (\ref{eq:theta-def}). The circuit consists of sequences of controlled $y$-rotation gates. Specifically, qubit $q_m$, where $0\le m< n$, includes $2^m$ such gates where the single-qubit $y$ rotations are of the form $R_y(\theta_{n,n-m}(i))$, where $i=0,1,\dots,2^m-1$. The open (closed) vertices represent negative (positive) controls, as determined from the base-2 representation of integer-argument $i$ of a given rotation.}
\end{figure*}

Similarly, in a second step, we can reduce the expression of $\ket{\psi_n}$ of (\ref{eq:psi}) to a form amenable to quantum computation. Specifically, making use of (\ref{eq:xi}), initially we have
\begin{equation}	\label{eq:psi-1}
\ket{\psi_n} = \sqrt{\delta x_n} \sum_{i_n} \sqrt{ \sum_{j=-\infty}^\infty P(x_n(i_n)+j w) } \cdot \ket{i_n} .
\end{equation}
For the specific case of $n=1$, we can leverage (\ref{eq:deltax-1}) and write
\begin{multline}	\label{eq:psi1}
\ket{\psi_1} = \sqrt{\frac{\sum_{j=-\infty}^\infty P(x_0+j w)}{\sum_{j=-\infty}^\infty P(x_0+j w/2)} } \cdot \ket{0} \\
+ \sqrt{\frac{\sum_{j=-\infty}^\infty P(x_0+\left(2j+1\right) w/2)}{\sum_{j=-\infty}^\infty P(x_0+j w/2)} } \cdot \ket{1} .
\end{multline}
The summation identity of (\ref{eq:sumIdentity}) allows us to introduce angles of rotation, $\theta_{n,m}(i) $, via the definitions
\begin{equation}	\label{eq:theta-1}
\cos{\frac{\theta_{n,m}(i)}{2}} = \sqrt{\frac{\sum_{j=-\infty}^\infty P(x_n(i)+j w/2^{m-1})}{\sum_{j=-\infty}^\infty P(x_n(i)+j w/2^m)}} , 
\end{equation}
\begin{equation}	\label{eq:theta-2}
\sin{\frac{\theta_{n,m}(i)}{2}} = \sqrt{\frac{\sum_{j=-\infty}^\infty P(x_n(i)+\left(2j+1\right) w/2^m)}{\sum_{j=-\infty}^\infty P(x_n(i)+j w/2^m)}} ,
\end{equation}
such that for $n=1$ and $m=1$ we have
\begin{equation}	\label{eq:psi1-2}
\ket{\psi_1} = \cos{\frac{\theta_{1,1}(0)}{2}} \cdot \ket{0} + \sin{\frac{\theta_{1,1}(0)}{2}} \cdot \ket{1} .
\end{equation}

More generally, for the case $n>1$, an initial divide of the sum over $0\le i_n<2^n$ into two parts, based on the value of the most significant bit, allows us to write (\ref{eq:psi-1}) as
\begin{multline}	\label{eq:psi-2}
\ket{\psi_n} = \sqrt{\delta x_n} \sum_{i_{n-1}} 
\Bigg( \sqrt{ \sum_{j=-\infty}^\infty P(x_n(i_{n-1})+j w) } \cdot \ket{0} \\
+ \sqrt{ \sum_{j=-\infty}^\infty P(x_n(i_{n-1})+\left(2j+1\right) w/2) } \cdot \ket{1} \Bigg) \otimes \ket{i_{n-1}}  .
\end{multline}
Then, making use of (\ref{eq:sumIdentity}), with the aid of (\ref{eq:theta-1}) and (\ref{eq:theta-2}), equation (\ref{eq:psi-2}) can be expressed as
\begin{multline}	\label{eq:psi-3}
\ket{\psi_n} = \sqrt{\delta x_n} \sum_{i_{n-1}} 
\Big( \cos{\frac{\theta_{n,1}(i_{n-1})}{2}} \cdot \ket{0} \\
+ \sin{\frac{\theta_{n,1}(i_{n-1})}{2}} \cdot \ket{1} \Big) \\
\otimes \sqrt{ \sum_{j=-\infty}^\infty P(x_n(i_{n-1})+j w/2) } \cdot \ket{i_{n-1}} .
\end{multline}
Again, dividing the sum into two parts with respect to the most significant bit, and making use of the three equations (\ref{eq:sumIdentity}), (\ref{eq:theta-1}) and (\ref{eq:theta-2}), yields
\begin{multline}	\label{eq:psi-4}
\ket{\psi_n} = \sqrt{\delta x_n} \sum_{i_{n-2}} \Bigg[
\Big( \cos{\frac{\theta_{n,1}(i_{n-2})}{2}} \cdot \ket{0} \\
+ \sin{\frac{\theta_{n,1}(i_{n-2})}{2}} \cdot \ket{1} \Big)  \otimes \cos{\frac{\theta_{n,2}(i_{n-2})}{2}} \cdot \ket{0} \\
+ \Big( \cos{\frac{\theta_{n,1}(i_{n-2}+2^{n-2})}{2}} \cdot \ket{0} + \sin{\frac{\theta_{n,1}(i_{n-2}+2^{n-2})}{2}} \cdot \ket{1} \Big) \\
\otimes \sin{\frac{\theta_{n,2}(i_{n-2})}{2}} \cdot \ket{1}  
\Bigg] \\
\otimes \sqrt{ \sum_{j=-\infty}^\infty P(x_n(i_{n-2})+j w/4) } \cdot \ket{i_{n-2}} .
\end{multline}

Repeating these steps, for a total of $n-1$ iterations, and applying (\ref{eq:deltax-1}) in the last iteration, yields a result that can be expressed as a sum of Kronecker products, viz.
\begin{equation}	\label{eq:psi-derived}
\ket{\psi_n} = \sum_{i_{n}} \bigotimes_{m=1}^n \cos{ \left( -\frac{\pi}{2} i^{(n)}_{n-m} + \frac{1}{2} \theta_{n,m}(i_{n-m}) \right)} \ket{i^{(n)}_{n-m}} ,
\end{equation}
where we recall the notation defined in (\ref{eq:integer}). Also, the definition in (\ref{eq:theta-1}) implies
\begin{multline}	\label{eq:theta-def}
\theta_{n,m}(i_{n-1}) = 2 \\
\times \arccos{ \sqrt{\frac{\sum_{j=-\infty}^\infty P(x_n(i_{n-1})+j w/2^{m-1})}{\sum_{j=-\infty}^\infty P(x_n(i_{n-1})+j w/2^m)}} } ,
\end{multline}
where we have replaced $i$ by $i_{n-1}$ since $0\le i < 2^{n-1}$. The $n$-qubit circuit diagram corresponding to (\ref{eq:psi-derived}) is depicted in Fig.~\ref{fig2} in terms of $y$-rotation gates parameterized by rotation angles $\theta_{n,m}(i)$, where $m=1,2,\dots,n$. The algorithm represented by the circuit of Fig.~\ref{fig2} is a form of divide-and-conquer algorithm \cite{Mottonen2005}.

\subsection{The Application to Discrete Probabilities\label{subsec:discrete}}
The quantum circuit of Fig.~\ref{fig2} can be leveraged for preparation of state vectors whose amplitude coefficients are the square roots of discrete probabilities. Specifically, consider a discrete distribution of probabilities of the form
\begin{equation}	\label{eq:discrete-p}
\mathcal{P}(i_n) = \frac{ f(i_n) }{\sum_{i_n} f(i_n) } \; ; \quad 
\sum_{i_n} \mathcal{P}(i_n) = 1 ,
\end{equation}
where $f(i_n)$ is a real number dependent on integer $i_n$, with corresponding quantum state vector
\begin{equation}	\label{eq:phi-M}
\ket{\phi_n} = \sum_{i_n} \sqrt{\mathcal{P}(i_n)} \, \ket{i_n} \; ; \quad \braket{\phi_n|\phi_n}=1 .
\end{equation}
It is not possible to prepare (\ref{eq:phi-M}) from the set of unitary gate operations depicted in Fig.~\ref{fig2}, except when $n=1$. In this special case, because $\mathcal{P}(1)$ is the compliment of $\mathcal{P}(0)$, we have
\begin{equation}	\label{eq:phi-1}
\ket{\phi_1} = \sqrt{\mathcal{P}(0)} \, \ket{0} + \sqrt{1-\mathcal{P}(0)} \, \ket{1} ,
\end{equation}
such that we can define
\begin{equation}
\theta_{1,1}(0) = 2 \arccos{ \sqrt{ \mathcal{P}(0) } } ,
\end{equation}
\begin{equation}
\ket{\psi_1} = \cos{\frac{\theta_{1,1}(0)}{2}} \cdot \ket{0} + \sin{\frac{\theta_{1,1}(0)}{2}} \cdot \ket{1} = \ket{\phi_1} .
\end{equation}

The form of (\ref{eq:phi-1}) suggests introducing an auxiliary state vector defined from complimentary probabilities, viz.
\begin{equation}	\label{eq:phi-c}
\ket{\phi^*_n} =\sum_{i_n} \sqrt{1-\mathcal{P}(i_n)} \, \ket{i_n} \; ; \quad \braket{\phi^*_n|\phi^*_n}=2^n-1 ,
\end{equation}
then defining the actual state vector of preparation as
\begin{equation}	\label{eq:psi-N+1}
\ket{\psi_n} = \frac{1}{\sqrt{2^{n-1}}} \left( \ket{0} \otimes \ket{\phi_{n-1}} + \ket{1} \otimes \ket{\phi^*_{n-1}} \right) \; ; \quad n > 1 .
\end{equation}
Now we have a state vector in (\ref{eq:psi-N+1}) that can be prepared via the quantum circuit of Fig.~\ref{fig2}. To see this, we first note that it has a corresponding PDF given by
\begin{multline}	\label{eq:pdf-discrete}
P(x) = \frac{1}{2^{n-1}} \sum_{i_{n-1}} \Bigg[ \mathcal{P}(i_{n-1}) \, \delta(x-x_n(i_{n-1})) \\
+ \left(1-\mathcal{P}(i_{n-1})\right) \delta(x-x_n(2^{n-1}+i_{n-1}))  \Bigg] \; ; \quad n > 1,
\end{multline}
as shown in Appendix~\ref{app:discrete}. Here, we have reused the definition of (\ref{eq:xmap}) for $x_n(i)$, but with $\bar{x}=0$, $\zeta=0$, and $w$ of arbitrary length. Also, we introduced the Dirac delta function, $\delta(x)$, noting
that (i) $P(x)=0$ when $|x|>w/2$ by construction, such that (\ref{eq:pdf-condition}) is satisfied exactly rather than approximately, i.e.,
\begin{equation}
P(x_n+j w) = P(x_n) \, \delta_{j,0} ,
\end{equation}
and (ii) $P(x)$ is normalized to unity, when integration is over all $x\in\R$. 

Parameters $\theta_{n,m}(i)$ are computed from (\ref{eq:theta-def}) using (\ref{eq:pdf-discrete}). In this case, as in Appendix~\ref{app:discrete}, the Dirac delta functions of (\ref{eq:pdf-discrete}) are replaced with Gaussians, i.e.,
\begin{equation}	\label{eq:dirac-gaussian}
\delta(x) = \lim_{\epsilon \rightarrow 0} \frac{1}{\sqrt{2 \pi \epsilon^2}} e^{-x^2 / 2 \epsilon^2} .
\end{equation}
As shown in Appendix~\ref{app:angle}, we evaluate sums over $j$ as $\epsilon\rightarrow 0$ in (\ref{eq:theta-def}) and obtain 
\begin{equation}	\label{eq:theta-discrete-0}
\theta_{n,m}(i_{n-1}) = \left\{
\begin{array}{ccl}
2 \arccos{ \sqrt{ \mathcal{P}(i_{n-1}) } } & ; & m = 1 \\
\pi / 2 & ; & 1 < m \le n
\end{array}
\right. .
\end{equation}

If (\ref{eq:theta-discrete-0}) is substituted into (\ref{eq:psi-derived}) then we arrive at an expression specific to discrete probability distributions given by
\begin{equation}	\label{eq:psi-discrete}
\ket{\psi_n} = \frac{1}{\sqrt{2^{n-1}}} \sum_{i_{n}} \cos{ \left( -\frac{\pi}{2} i^{(n)}_{n-1} + \frac{\theta(i_{n-1})}{2} \right)} \ket{i_n} ,
\end{equation}
where we have simplified notation to include only the angles of the $y$-rotation gates corresponding to $m=1$, viz.
\begin{equation}	\label{eq:theta-discrete}
\theta(i_{n-1}) = 2 \arccos{ \sqrt{ \mathcal{P}(i_{n-1}) } } .
\end{equation}

\begin{figure}[h]
\center
\includegraphics[width=240pt, height=140pt]{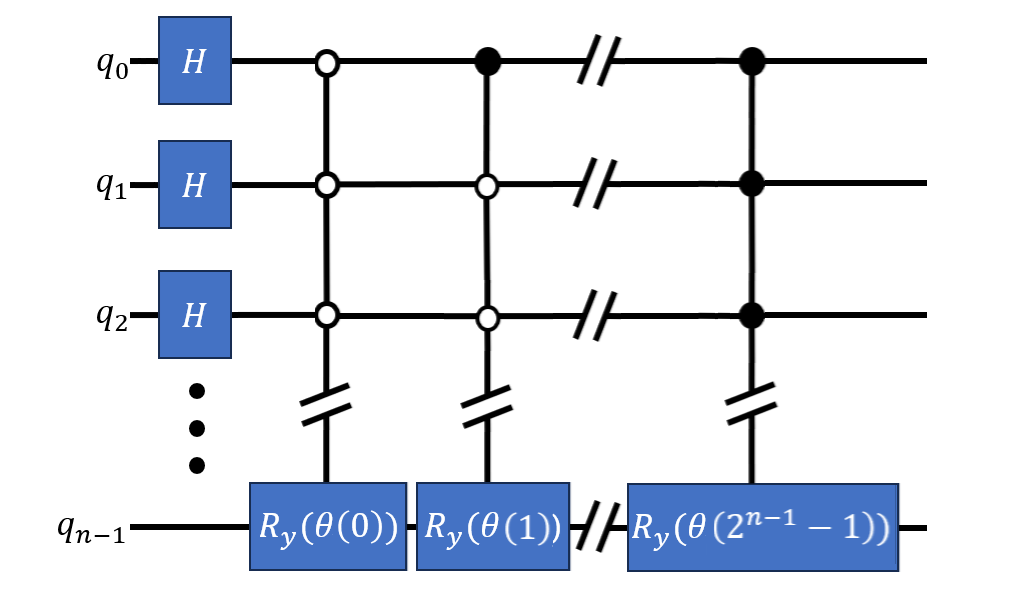}
\caption{\label{fig3} The $n$-qubit circuit diagram of state vector $\ket{\psi_n}$, of (\ref{eq:psi-N+1}), parameterized by rotation angles $\theta(i)$, as defined in (\ref{eq:theta-discrete}). Hadamard gates, $H$, are applied to the lesser significant qubits, $q_0, q_1, \dots, q_{n-2}$, while controlled y-rotations, by angle $\theta(i)$, are applied thereafter, in sequence, with $i=0,1,\dots,2^{n-1}-1$. The open (closed) vertices represent negative (positive) controls, as determined from the base-2 representation of integer-argument $i$ of a given rotation.}
\end{figure}

Figure~\ref{fig3} shows the circuit diagram for preparing the state vector of (\ref{eq:psi-discrete}); gates corresponding to $\pi/2$ rotation angle reduce to Hadamard gates, $H$, and even multiples of these gates reduce further to identity operations. It is easy to see that (\ref{eq:psi-N+1}) and (\ref{eq:psi-discrete}) are both expressions of the same state vector, $\ket{\psi_n}$. For example, comparing the two expressions, we have
\begin{equation}
\ket{\phi_n} = \sum_{i_n} \cos{ \left( \frac{\theta(i_{n})}{2} \right)} \ket{i_n} = \sum_{i_n} \sqrt{\mathcal{P}(i_n)} \ket{i_n} .
\end{equation}

\subsection{The Quantum Forking Algorithm}
We have shown how to construct a divide-and-conquer upsampling algorithm for quantum state preparation. The quantum circuit corresponding to this algorithm, as depicted in Fig.~\ref{fig2}, exhibits exponential gate depth, but can be transformed to an equivalent circuit of polylogarthmic gate depth, as described by Araujo and coworkers \cite{Araujo2021}, using a method of quantum forking \cite{Park2019b}. The source code of Araujo and coworkers can be found on the GitHub site \cite{daSilvaCode2020}.

Compared to the $n$-qubit divide-and-conquer upsampling algorithm, the method of \cite{Araujo2021} introduces a control qubit, $d=2^n-1$ ancilla qubits, and a set of $d-1$ controlled swap gates, as depicted in Fig.~1 of \cite{Park2019b}, reducing time complexity by a factor of $d$, with a gate depth of $O(\log_2^2(d))$. In the examples that follow, we convert the upsampling algorithm to the quantum-forking algorithm, explicitly showing the quantum-forking circuit used in our calculations.

\section{Examples\label{sec:examples}}
We present simulations of state-vector preparation, for both continuous and discrete probability examples. Continuous-probability examples include state vectors of Gaussian and Laplace distributions while a discrete-probability example shown is that of the binomial distribution. We include illustrations of the specific quantum-forking quantum circuits used in our calculations.
\begin{figure}[ht]
\center
\includegraphics[width=240pt, height=280pt]{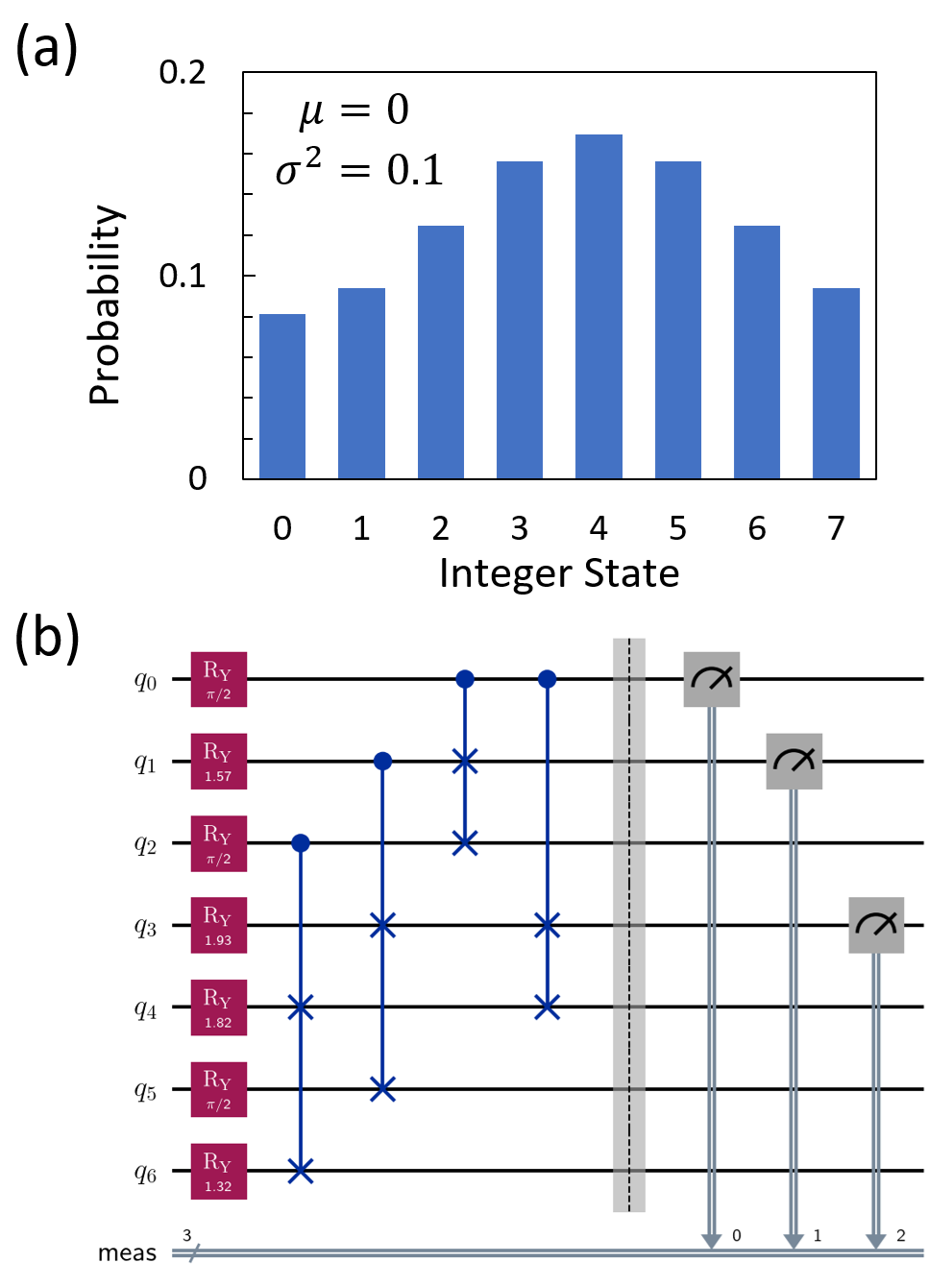}
\caption{\label{fig4} (a) Gaussian distributed probabilities corresponding to a state vector prepared via the quantum-forking circuit of panel (b). The original upsampling circuit, prior to conversion to the quantum-forking circuit, consists of $n=3$ qubits.}
\end{figure}

\subsection{Continuous Probability Distributions\label{subsec:continuous}}

\subsubsection{The Gaussian Distribution}
The univariate normal, or Gaussian, probability density can be expressed as
\begin{equation}	\label{eq:pdfnormal}
P(x|\mu,\sigma) = \frac{1}{\sigma\sqrt{2\pi}} \, \exp{ \left[ - \frac{ { \left( x - \mu \right) }^2 }{ 2 \sigma^2} \right] } ,
\end{equation}
where the upsampling center of capture on the line of support is $\bar{x}=\mu$, with
\begin{equation}
\mu = \int^\infty_{-\infty} x P(x|\mu,\sigma) \, dx ,
\end{equation}
\begin{equation}
\sigma^2 = \int^\infty_{-\infty} { \left( x - \mu \right) }^2 P(x|\mu,\sigma) \, dx .
\end{equation}
Figure~\ref{fig4}(a) shows the sampling measurement of a state vector prepared using the quantum-forking circuit of Fig.~\ref{fig4}(b). The quantum-forking circuit was the result of converting an upsampling circuit of $n=3$ qubits.

\subsubsection{The Laplace Distribution}
The Laplace probability distribution can be expressed as
\begin{equation}	\label{eq:pdflaplace}
P(x|\mu,b) = \frac{1}{2 b} \, \exp{ \left( - \frac{ \left| x - \mu \right| }{b} \right) } ,
\end{equation}
where the upsampling center of capture on the line of support is $\bar{x}=\mu$, with
\begin{equation}
\mu = \int^\infty_{-\infty} x P(x|\mu,b) \, dx ,
\end{equation}
\begin{equation}
b^2 = \frac{1}{2} \int^\infty_{-\infty} { \left( x - \mu \right) }^2 P(x|\mu,b) \, dx .
\end{equation}
\begin{figure}[ht]
\center
\includegraphics[width=240pt, height=280pt]{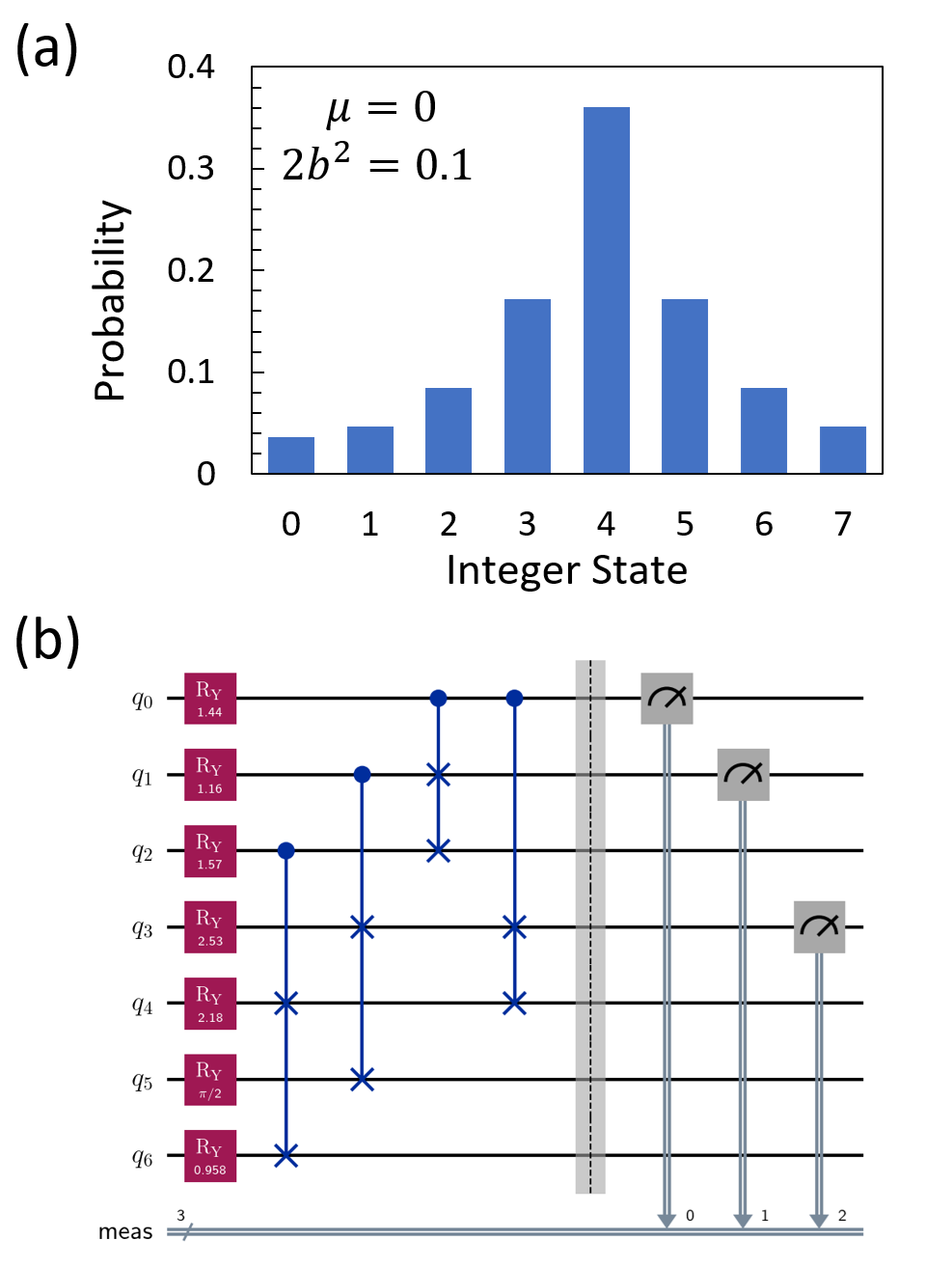}
\caption{\label{fig5} (a) Laplace distributed probabilities corresponding to a state vector prepared via the quantum-forking circuit of panel (b). The original upsampling circuit, prior to conversion to the quantum-forking circuit, consists of $n=3$ qubits.}
\end{figure}
Figure~\ref{fig5}(a) shows the sampling measurement of a state vector prepared using the quantum-forking circuit of Fig.~\ref{fig5}(b). As in Fig.~\ref{fig4}, the quantum-forking circuit was the result of converting an upsampling circuit of $n=3$ qubits.

\subsubsection{The Lognormal Distribution}
Also, the upsampling algorithm can be applied to continuous probability distributions that can be mapped to $x\in\R$. An example is the lognormal probability distribution, given by
\begin{equation}	\label{eq:pdflognormal}
P(y|\mu,\sigma) = \frac{1}{y\sigma\sqrt{2\pi}} \, \exp{ \left[ - \frac{ { \left( \log{y} - \mu \right) }^2 }{ 2 \sigma^2} \right] } .
\end{equation}
Here, the mapping is $x=\log{y}$ such that $P(y|\mu,\sigma) \, dy = P(x|\mu,\sigma) \, dx$, where $P(x|\mu,\sigma)$ is the Gaussian distribution of (\ref{eq:pdfnormal}). The discrete set of points sampled on the interval of $0\le y < \infty$ are defined via $y_n=e^{x_n}$, and the width of the probability at $y_n$, i.e., related to the area under the curve, is $\delta y_n=e^{x_n} \delta x_n$.

\subsection{Discrete Probability Distributions\label{subsec:discrete}}

\subsubsection{The Binomial Distribution}
As an example of a discrete case we consider the binomial distribution. Here, we define binomial probabilities in the form
\begin{equation}	\label{eq:pdfbinomial}
\mathcal{P}(k,l) = \left( 
\begin{array}{c}
l \\
k
\end{array}
\right) p^k q^{l-k} \; ; \quad k \in \left\{ 0, 1, 2, \dots, l \right\} ,
\end{equation}
where the support is the set of integers $k$, such that $0 \le k \le l$, and we define probabilities $0 \le p,q\le 1$, where $p+q=1$.
\begin{figure}[h]
\center
\includegraphics[width=240pt, height=300pt]{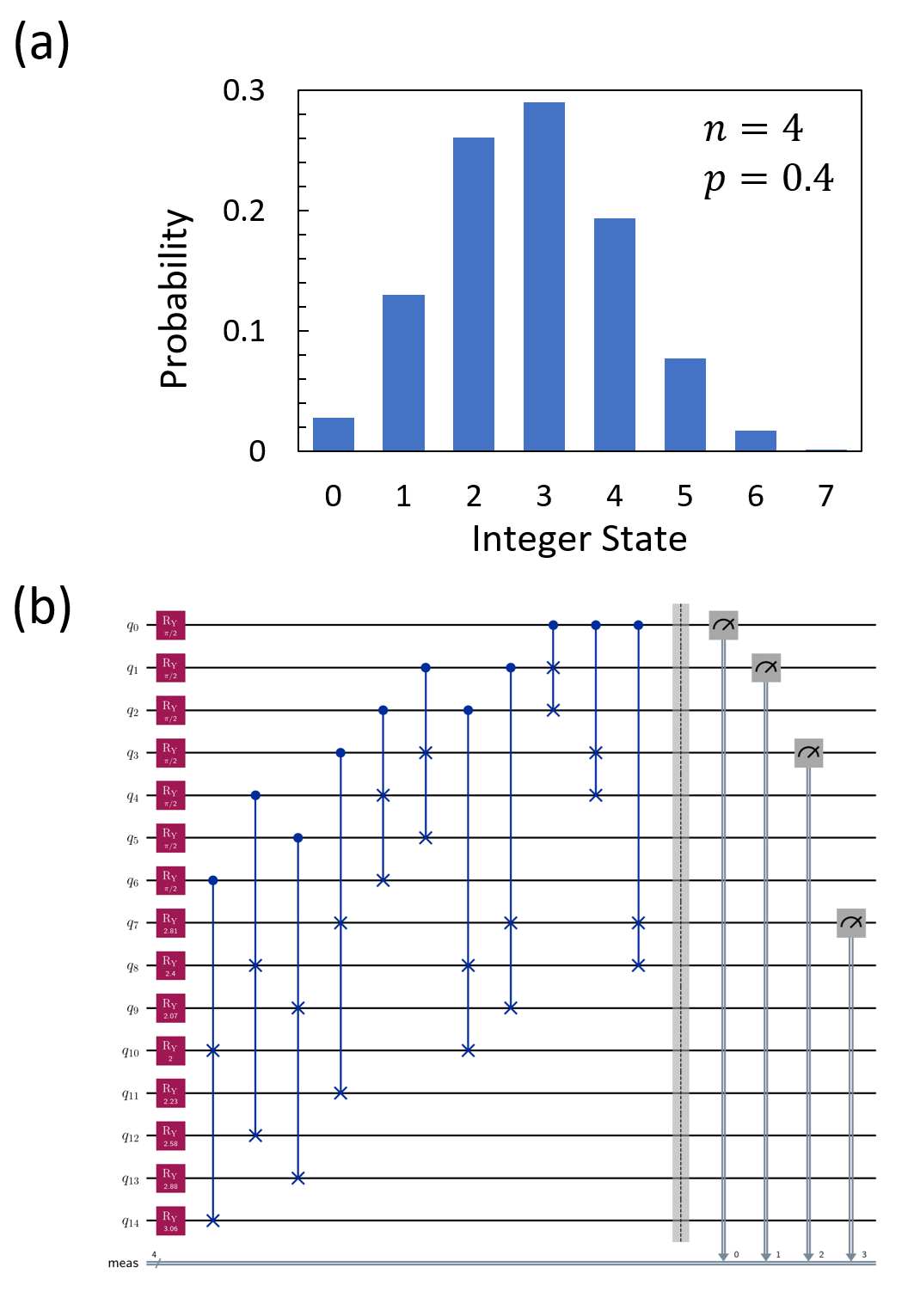}
\caption{\label{fig6} (a) Plot of the binomial distribution obtained from projection of the $\ket{0}$ state of the most significant qubit of a $n=4$ quantum circuit converted to the quantum-forking circuit depicted in panel (b).}
\end{figure}

\section{Conclusion\label{sec:conclusion}}
We have presented the upsampling algorithm of state-vector preparation, applicable to both continuous probability distributions with support on $x\in\R$, as well as arbitrary discrete probability distributions. The upsampling algorithm is a divide-and-conquer algorithm exhibiting exponential gate depth but it can be transformed to a quantum-forking algorithm of polylogarithmic gate depth \cite{Araujo2021,Park2019b,daSilvaCode2020}. We presented several examples, including the Gaussian, Laplace, and binomial distributions, and we showed the quantum-forking circuits we employed for our simulations.

Our analyses focused on univariate distributions, but our approach can be extended to multivariate forms. Also, the algorithm is adaptable to the preparation of state vectors associated with other functions. We hope our discussion will stimulate further interest and development.

\begin{acknowledgments}
We thank V. Markov and C. Gonciulea for helpful comments and suggestions, with additional support and encouragement from E. Almond, T. Sewell, and T. Skeen. The views expressed in this article are those of the authors and do not represent the views of Wells Fargo. This article is for informational purposes only. Nothing contained in this article should be construed as investment advice. Wells Fargo makes no express or implied warranties and expressly disclaims all legal, tax, and accounting implications related to this article.
\end{acknowledgments}

\appendix

\section{The Upsampling Algorithm Applied to Discrete Probabilities\label{app:discrete}}
We can apply the upsampling algorithm to preparation of state vectors whose coefficients are the square roots of discrete probabilities. In this case, we want to show that (\ref{eq:pdf-discrete}) is the PDF corresponding to state vector $\ket{\psi_n}$, as written in the form of (\ref{eq:psi-N+1}). As a first step, we represent the Dirac delta function by a Gaussian whose variance approaches zero, as in (\ref{eq:dirac-gaussian}). In our calculations that utilize (\ref{eq:dirac-gaussian}) we omit the explicit notation of a limit, but it's understood that we must take the limit as $\epsilon\rightarrow 0$ at the end of all calculations. 

Next, assuming $n>1$, we apply (\ref{eq:pdf-discrete}) and (\ref{eq:dirac-gaussian}) to (\ref{eq:xi}) to obtain an expression for amplitude coefficients, $\xi_n(i_n)$, of $\ket{\psi_n}$, as expressed in the form of (\ref{eq:psi}). Introducing integer $k_n$, of the same form as $i_n$, where $0\le k_n<2^n$, the amplitude $\xi_n(i_n)$ can be expressed as
\begin{widetext}
\begin{multline}	\label{eq:xi-discrete}
\xi_n(i_n) = \Bigg\{ \frac{\delta x_n}{2^{n-1} \sqrt{2 \pi \epsilon^2}}
\sum_{k_{n-1}} \Bigg[ \mathcal{P}(k_{n-1}) \sum_{j=-\infty}^\infty  e^{-\left( w^2 / 8\epsilon^2 \right) { \left[ 2j + \left( i_n - k_{n-1} \right) / 2^{n-1} \right] }^2 } \\
+ \left( 1-\mathcal{P}(k_{n-1}) \right) \sum_{j=-\infty}^\infty  e^{-\left( w^2 / 8\epsilon^2 \right)
{ \left[ 2j - 1 + \left( i_n - k_{n-1} \right) / 2^{n-1} \right] }^2 } \Bigg] \Bigg\} ^{1/2} \; ; \quad n > 1 .
\end{multline}
To evaluate normalization factor $\delta x_n$, we first note that $\braket{\psi_n|\psi_n}=1$ implies
\begin{equation}
\sum_{i_n} \xi_n(i_n)^2 = 1 ,
\end{equation}
such that, inserting (\ref{eq:xi-discrete}) into this expression, then solving for $\delta x_n$, yields
\begin{multline}	\label{eq:norm-discrete}
\delta x_n = 2^{n-1} \sqrt{2 \pi \epsilon^2} \Bigg\{ \sum_{i_n} \sum_{k_{n-1}} 
\Bigg[ \mathcal{P}(k_{n-1}) \sum_{j=-\infty}^\infty  e^{-\left( w^2 / 8\epsilon^2 \right) { \left[ 2j + \left( i_n - k_{n-1} \right) / 2^{n-1} \right] }^2 } \\
+ \left( 1-\mathcal{P}(k_{n-1}) \right) \sum_{j=-\infty}^\infty  e^{-\left( w^2 / 8\epsilon^2 \right)
{ \left[ 2j - 1 + \left( i_n - k_{n-1} \right) / 2^{n-1} \right] }^2 } \Bigg] \Bigg\} ^{-1} .
\end{multline}
Substituting (\ref{eq:norm-discrete}) into (\ref{eq:xi-discrete}) gives
\begin{multline}	\label{eq:xi-discrete-2}
\xi_n(i_n) = \Bigg(
\sum_{k_{n-1}} \Bigg[ \mathcal{P}(k_{n-1}) \sum_{j=-\infty}^\infty  e^{-\left( w^2 / 8\epsilon^2 \right) { \left[ 2j + \left( i_n - k_{n-1} \right) / 2^{n-1} \right] }^2 } \\
+ \left( 1-\mathcal{P}(k_{n-1}) \right) \sum_{j=-\infty}^\infty  e^{-\left( w^2 / 8\epsilon^2 \right)
{ \left[ 2j - 1 + \left( i_n - k_{n-1} \right) / 2^{n-1} \right] }^2 } \Bigg] \\
\times \Bigg\{ \sum_{i_n} \sum_{k_{n-1}} 
\Bigg[ \mathcal{P}(k_{n-1}) \sum_{j=-\infty}^\infty  e^{-\left( w^2 / 8\epsilon^2 \right) { \left[ 2j + \left( i_n - k_{n-1} \right) / 2^{n-1} \right] }^2 } \\
+ \left( 1-\mathcal{P}(k_{n-1}) \right) \sum_{j=-\infty}^\infty  e^{-\left( w^2 / 8\epsilon^2 \right)
{ \left[ 2j - 1 + \left( i_n - k_{n-1} \right) / 2^{n-1} \right] }^2 } \Bigg] \Bigg\} ^{-1} \Bigg) ^{1/2} \; ; \quad n > 1 .
\end{multline}

As $\epsilon\rightarrow 0$ on the right side of (\ref{eq:xi-discrete-2}), most of the terms in the sums over $j$ vanish. The exceptional terms, which approach unity, have exponential functions with arguments that go to zero. The zero condition only occurs if the difference $i_n - k_{n-1}$ is divisible by $2^{n-1}$. One can show that only $j=0$ terms are non-zero, and only when $i_n=k_{n-1}$ or $i_n=2^{n-1}+k_{n-1}$. In particular, when $i_n=k_{n-1}$ then the most significant bit of $i_n$ is zero, i.e., $i^{(n)}_{n-1}=0$, such that $i_n=i_{n-1}=k_{n-1}$; when $i_n=2^{n-1}+k_{n-1}$ then the most significant bit of $i_n$ is unity, i.e., $i^{(n)}_{n-1}=1$, such that  $i_n=2^{n-1}+i_{n-1}$, or $i_{n-1}=k_{n-1}$. 

Taking these observations into account, then as $\epsilon\rightarrow 0$ in (\ref{eq:xi-discrete-2}) we obtain an intermediate result
\begin{multline}	\label{eq:xi-discrete-3}
\xi_n(i_n) = \Bigg( \Bigg[ \mathcal{P}(i_{n-1}) \, \delta_{i^{(n)}_{n-1},0} + \left( 1-\mathcal{P}(i_{n-1}) \right) \delta_{i^{(n)}_{n-1},1}  \Bigg] \\
\times \Bigg\{ \sum_{i_n} \Bigg[ \mathcal{P}(i_{n-1}) \, \delta_{i^{(n)}_{n-1},0} + \left( 1-\mathcal{P}(i_{n-1}) \right) \, \delta_{i^{(n)}_{n-1},1} \Bigg] 
\Bigg\} ^{-1} \Bigg) ^{1/2} \; ; \quad n > 1 .
\end{multline}
But, making use of the normalization condition in (\ref{eq:discrete-p}), we have
\begin{equation}
\sum_{i_n} \Bigg[ \mathcal{P}(i_{n-1}) \, \delta_{i^{(n)}_{n-1},0} + \left( 1-\mathcal{P}(i_{n-1}) \right) \, \delta_{i^{(n)}_{n-1},1} \Bigg] =
\sum_{i_{n-1}} \Bigg[ \mathcal{P}(i_{n-1}) + \left( 1-\mathcal{P}(i_{n-1}) \right) \Bigg] = 2^{n-1} ,
\end{equation}
so we can reduce the expression of $\xi_n(i_n)$ to
\begin{equation}	\label{eq:xi-discrete-4}
\xi_n(i_n) = \sqrt{ \frac{1}{2^{n-1}} \Bigg[ \mathcal{P}(i_{n-1}) \, \delta_{i^{(n)}_{n-1},0} + \left( 1-\mathcal{P}(i_{n-1}) \right) \delta_{i^{(n)}_{n-1},1}  \Bigg] } \; ; \quad n > 1 .
\end{equation}
Substituting (\ref{eq:xi-discrete-4}) into (\ref{eq:psi}) yields an expression for $\ket{\psi_n}$ identical to (\ref{eq:psi-N+1}). This shows that (\ref{eq:pdf-discrete}) is the PDF corresponding to (\ref{eq:psi-N+1}).

\section{The Angle Parameters of the Discrete-Probability Algorithm\label{app:angle}}
The $y$-rotation angles, $\theta_{n,m}(i)$, of the upsampling algorithm, applicable to a state vector prepared from discrete probabilities of the form of (\ref{eq:discrete-p}), can be computed from (\ref{eq:theta-def}) using (\ref{eq:pdf-discrete}). As in Appendix~\ref{app:discrete}, we replace the Dirac delta functions of (\ref{eq:pdf-discrete}) with Gaussian functions, via (\ref{eq:dirac-gaussian}). Expressing the finite sums as sums over integers $k_{n-1}$, where $0\le k_{n-1}<2^{n-1}$, we then have
\begin{multline}	\label{eq:theta-def-0}
\theta_{n,m}(i_{n-1}) = 2 \arccos{}
\Bigg\{ \sum_{k_{n-1}} \Bigg[ 
\mathcal{P}(k_{n-1}) \sum_{j=-\infty}^\infty e^{-\left[ w^2 / \left( 2^{2m+1} \epsilon^2 \right) \right] {\left[ 2 j + \left( i_{n-1} - k_{n-1} \right) / 2^{n-m} \right]}^2} \\
+ \left(1-\mathcal{P}(k_{n-1})\right) \sum_{j=-\infty}^\infty e^{-\left[ w^2 / \left( 2^{2m+1} \epsilon^2 \right) \right] {\left[ 2 j - 2^{m-1} + \left( i_{n-1} - k_{n-1} \right) / 2^{n-m} \right]}^2} 
\Bigg] \Bigg\} ^{1/2} \\
\times\Bigg\{ \sum_{k_{n-1}} \Bigg[ 
\mathcal{P}(k_{n-1}) \sum_{j=-\infty}^\infty e^{-\left[ w^2 / \left( 2^{2m+1} \epsilon^2 \right) \right] {\left[ j + \left( i_{n-1} - k_{n-1} \right) / 2^{n-m} \right]}^2} \\
+ \left(1-\mathcal{P}(k_{n-1})\right) \sum_{j=-\infty}^\infty e^{-\left[ w^2 / \left( 2^{2m+1} \epsilon^2 \right) \right] {\left[ j - 2^{m-1} + \left( i_{n-1} - k_{n-1} \right) / 2^{n-m} \right]}^2} 
\Bigg] \Bigg\} ^{-1/2} ,
\end{multline}
\end{widetext}
where we removed explicit use of limits, with the understanding that at the end of calculation we will let $\epsilon\rightarrow 0$.

As in Appendix~\ref{app:discrete}, when $\epsilon\rightarrow 0$ on the right side of (\ref{eq:theta-def-0}), most of the terms in the sums over $j$ vanish. The exceptional terms, which approach unity, have exponential functions with arguments that go to zero. The zero argument only occurs if the difference $i_{n-1} - k_{n-1}$ is divisible by $2^{n-m}$, leading to subsets of values of $k_{n-1}$ of the form $k_{n-1}=2^{n-m} \cdot l + i_{n-1}$, corresponding to one or more integers $l\ge 0$.

When $m=1$, the first sum over $j$ in the numerator, proportional to $\mathcal{P}(k_{n-1})$, yields the constraint $k_{n-1}=i_{n-1}$, such that the sum over $k_{n-1}$ reduces to the single term $\mathcal{P}(i_{n-1})$. On the other hand, the second sum over $j$ of the numerator, proportional to $\left( 1 - \mathcal{P}(k_{n-1}) \right)$, is precisely zero---no terms in this sum over $k_{n-1}$ contribute. In the denominator, when again $m=1$, the two sums over $j$ contribute equally, again producing the constraint $k_{n-1}=i_{n-1}$, which means we have a sum of $\mathcal{P}(i_{n-1})$ and $\left( 1 - \mathcal{P}(i_{n-1}) \right)$, which is unity, so the entire denominator evaluates to unity. Thus, in the $m=1$ case, as $\epsilon\rightarrow 0$, equation (\ref{eq:theta-def-0}) simplifies to
\begin{equation}	\label{eq:theta-m-1}
\theta_{n,1}(i_{n-1}) = 2 \arccos{ \sqrt{ \mathcal{P}(i_{n-1}) } } .
\end{equation}

When $m>1$, the two sums over $j$ of the numerator contribute equally, and in the denominator the two sums over $j$ also contribute equally. This means that in both numerator and denominator there are terms $\mathcal{P}(k_{n-1})$ and $\left( 1 - \mathcal{P}(k_{n-1}) \right)$, for values of $k_{n-1}$ of the form $k_{n-1}=2^{n-m} \cdot l + i_{n-1}$, that sum to unity. However, in the denominator, the count of such terms is twice that of the numerator, so in the case of $m>1$, as $\epsilon\rightarrow 0$, equation (\ref{eq:theta-def-0}) simplifies to
\begin{equation}	\label{eq:theta-m-greater}
\theta_{n,m}(i_{n-1}) = 2 \arccos{ \left( \frac{1}{\sqrt{2}} \right) } = \pi / 2 \; ; \quad m > 1 .
\end{equation}
Combining the two results, we can write the angle parameters of the discrete-probability algorithm as
\begin{equation}	\label{eq:theta-discrete-app}
\theta_{n,m}(i_{n-1}) = \left\{
\begin{array}{ccl}
2 \arccos{ \sqrt{ \mathcal{P}(i_{n-1}) } } & ; & m = 1 \\
\pi / 2 & ; & 1 < m \le n
\end{array}
\right. .
\end{equation}

\bibliography{bibliography}

%apsrev4-2.bst 2019-01-14 (MD) hand-edited version of apsrev4-1.bst
%Control: key (0)
%Control: author (8) initials jnrlst
%Control: editor formatted (1) identically to author
%Control: production of article title (-1) disabled
%Control: page (0) single
%Control: year (1) truncated
%Control: production of eprint (0) enabled
\begin{thebibliography}{25}%
\makeatletter
\providecommand \@ifxundefined [1]{%
 \@ifx{#1\undefined}
}%
\providecommand \@ifnum [1]{%
 \ifnum #1\expandafter \@firstoftwo
 \else \expandafter \@secondoftwo
 \fi
}%
\providecommand \@ifx [1]{%
 \ifx #1\expandafter \@firstoftwo
 \else \expandafter \@secondoftwo
 \fi
}%
\providecommand \natexlab [1]{#1}%
\providecommand \enquote  [1]{``#1''}%
\providecommand \bibnamefont  [1]{#1}%
\providecommand \bibfnamefont [1]{#1}%
\providecommand \citenamefont [1]{#1}%
\providecommand \href@noop [0]{\@secondoftwo}%
\providecommand \href [0]{\begingroup \@sanitize@url \@href}%
\providecommand \@href[1]{\@@startlink{#1}\@@href}%
\providecommand \@@href[1]{\endgroup#1\@@endlink}%
\providecommand \@sanitize@url [0]{\catcode `\\12\catcode `\$12\catcode
  `\&12\catcode `\#12\catcode `\^12\catcode `\_12\catcode `\%12\relax}%
\providecommand \@@startlink[1]{}%
\providecommand \@@endlink[0]{}%
\providecommand \url  [0]{\begingroup\@sanitize@url \@url }%
\providecommand \@url [1]{\endgroup\@href {#1}{\urlprefix }}%
\providecommand \urlprefix  [0]{URL }%
\providecommand \Eprint [0]{\href }%
\providecommand \doibase [0]{https://doi.org/}%
\providecommand \selectlanguage [0]{\@gobble}%
\providecommand \bibinfo  [0]{\@secondoftwo}%
\providecommand \bibfield  [0]{\@secondoftwo}%
\providecommand \translation [1]{[#1]}%
\providecommand \BibitemOpen [0]{}%
\providecommand \bibitemStop [0]{}%
\providecommand \bibitemNoStop [0]{.\EOS\space}%
\providecommand \EOS [0]{\spacefactor3000\relax}%
\providecommand \BibitemShut  [1]{\csname bibitem#1\endcsname}%
\let\auto@bib@innerbib\@empty
%</preamble>
\bibitem [{\citenamefont {Nielsen}\ and\ \citenamefont
  {Chuang}(2010)}]{Nielsen_Chuang_2010}%
  \BibitemOpen
  \bibfield  {author} {\bibinfo {author} {\bibfnamefont {M.~A.}\ \bibnamefont
  {Nielsen}}\ and\ \bibinfo {author} {\bibfnamefont {I.~L.}\ \bibnamefont
  {Chuang}},\ }\href@noop {} {\emph {\bibinfo {title} {Quantum Computation and
  Quantum Information: 10th Anniversary Edition}}}\ (\bibinfo  {publisher}
  {Cambridge University Press},\ \bibinfo {year} {2010})\ pp.\ \bibinfo {pages}
  {24--25}\BibitemShut {NoStop}%
\bibitem [{\citenamefont {Aaronson}(2015)}]{Aaronson2015}%
  \BibitemOpen
  \bibfield  {author} {\bibinfo {author} {\bibfnamefont {S.}~\bibnamefont
  {Aaronson}},\ }\href {https://doi.org/10.1038/nphys3272} {\bibfield
  {journal} {\bibinfo  {journal} {Nat. Phys.}\ }\textbf {\bibinfo {volume}
  {11}},\ \bibinfo {pages} {291} (\bibinfo {year} {2015})}\BibitemShut
  {NoStop}%
\bibitem [{\citenamefont {Biamonte}\ \emph {et~al.}(2017)\citenamefont
  {Biamonte}, \citenamefont {Wittek}, \citenamefont {Pancotti}, \citenamefont
  {Rebentrost}, \citenamefont {Wiebe},\ and\ \citenamefont
  {Lloyd}}]{Biamonte2017}%
  \BibitemOpen
  \bibfield  {author} {\bibinfo {author} {\bibfnamefont {J.}~\bibnamefont
  {Biamonte}}, \bibinfo {author} {\bibfnamefont {P.}~\bibnamefont {Wittek}},
  \bibinfo {author} {\bibfnamefont {N.}~\bibnamefont {Pancotti}}, \bibinfo
  {author} {\bibfnamefont {P.}~\bibnamefont {Rebentrost}}, \bibinfo {author}
  {\bibfnamefont {N.}~\bibnamefont {Wiebe}},\ and\ \bibinfo {author}
  {\bibfnamefont {S.}~\bibnamefont {Lloyd}},\ }\href
  {https://doi.org/10.1038/nature23474} {\bibfield  {journal} {\bibinfo
  {journal} {Nature}\ }\textbf {\bibinfo {volume} {549}},\ \bibinfo {pages}
  {195} (\bibinfo {year} {2017})}\BibitemShut {NoStop}%
\bibitem [{\citenamefont {Grover}\ and\ \citenamefont
  {Rudolph}(2002)}]{Grover2002}%
  \BibitemOpen
  \bibfield  {author} {\bibinfo {author} {\bibfnamefont {L.}~\bibnamefont
  {Grover}}\ and\ \bibinfo {author} {\bibfnamefont {T.}~\bibnamefont
  {Rudolph}},\ }\href {https://arxiv.org/abs/quant-ph/0208112} {\bibinfo
  {title} {Creating superpositions that correspond to efficiently integrable
  probability distributions}} (\bibinfo {year} {2002}),\ \Eprint
  {https://arxiv.org/abs/quant-ph/0208112} {arXiv:quant-ph/0208112 [quant-ph]}
  \BibitemShut {NoStop}%
\bibitem [{\citenamefont {Carrera~Vazquez}\ and\ \citenamefont
  {Woerner}(2021)}]{CarreraVazquez2021}%
  \BibitemOpen
  \bibfield  {author} {\bibinfo {author} {\bibfnamefont {A.}~\bibnamefont
  {Carrera~Vazquez}}\ and\ \bibinfo {author} {\bibfnamefont {S.}~\bibnamefont
  {Woerner}},\ }\href {https://doi.org/10.1103/PhysRevApplied.15.034027}
  {\bibfield  {journal} {\bibinfo  {journal} {Phys. Rev. Appl.}\ }\textbf
  {\bibinfo {volume} {15}},\ \bibinfo {pages} {034027} (\bibinfo {year}
  {2021})}\BibitemShut {NoStop}%
\bibitem [{\citenamefont {Zoufal}\ \emph {et~al.}(2019)\citenamefont {Zoufal},
  \citenamefont {Lucchi},\ and\ \citenamefont {Woerner}}]{Zoufal2019}%
  \BibitemOpen
  \bibfield  {author} {\bibinfo {author} {\bibfnamefont {C.}~\bibnamefont
  {Zoufal}}, \bibinfo {author} {\bibfnamefont {A.}~\bibnamefont {Lucchi}},\
  and\ \bibinfo {author} {\bibfnamefont {S.}~\bibnamefont {Woerner}},\ }\href
  {https://doi.org/10.1038/s41534-019-0223-2} {\bibfield  {journal} {\bibinfo
  {journal} {npj Quantum Inf.}\ }\textbf {\bibinfo {volume} {5}},\ \bibinfo
  {pages} {103} (\bibinfo {year} {2019})}\BibitemShut {NoStop}%
\bibitem [{\citenamefont {Zhu}\ \emph {et~al.}(2022)\citenamefont {Zhu},
  \citenamefont {Johri}, \citenamefont {Bacon}, \citenamefont {Esencan},
  \citenamefont {Kim}, \citenamefont {Muir}, \citenamefont {Murgai},
  \citenamefont {Nguyen}, \citenamefont {Pisenti}, \citenamefont {Schouela},
  \citenamefont {Sosnova},\ and\ \citenamefont {Wright}}]{Zhu2022}%
  \BibitemOpen
  \bibfield  {author} {\bibinfo {author} {\bibfnamefont {E.~Y.}\ \bibnamefont
  {Zhu}}, \bibinfo {author} {\bibfnamefont {S.}~\bibnamefont {Johri}}, \bibinfo
  {author} {\bibfnamefont {D.}~\bibnamefont {Bacon}}, \bibinfo {author}
  {\bibfnamefont {M.}~\bibnamefont {Esencan}}, \bibinfo {author} {\bibfnamefont
  {J.}~\bibnamefont {Kim}}, \bibinfo {author} {\bibfnamefont {M.}~\bibnamefont
  {Muir}}, \bibinfo {author} {\bibfnamefont {N.}~\bibnamefont {Murgai}},
  \bibinfo {author} {\bibfnamefont {J.}~\bibnamefont {Nguyen}}, \bibinfo
  {author} {\bibfnamefont {N.}~\bibnamefont {Pisenti}}, \bibinfo {author}
  {\bibfnamefont {A.}~\bibnamefont {Schouela}}, \bibinfo {author}
  {\bibfnamefont {K.}~\bibnamefont {Sosnova}},\ and\ \bibinfo {author}
  {\bibfnamefont {K.}~\bibnamefont {Wright}},\ }\href
  {https://doi.org/10.1103/PhysRevResearch.4.043092} {\bibfield  {journal}
  {\bibinfo  {journal} {Phys. Rev. Res.}\ }\textbf {\bibinfo {volume} {4}},\
  \bibinfo {pages} {043092} (\bibinfo {year} {2022})}\BibitemShut {NoStop}%
\bibitem [{\citenamefont {Sanders}\ \emph {et~al.}(2019)\citenamefont
  {Sanders}, \citenamefont {Low}, \citenamefont {Scherer},\ and\ \citenamefont
  {Berry}}]{Sanders2019}%
  \BibitemOpen
  \bibfield  {author} {\bibinfo {author} {\bibfnamefont {Y.~R.}\ \bibnamefont
  {Sanders}}, \bibinfo {author} {\bibfnamefont {G.~H.}\ \bibnamefont {Low}},
  \bibinfo {author} {\bibfnamefont {A.}~\bibnamefont {Scherer}},\ and\ \bibinfo
  {author} {\bibfnamefont {D.~W.}\ \bibnamefont {Berry}},\ }\href
  {https://doi.org/10.1103/PhysRevLett.122.020502} {\bibfield  {journal}
  {\bibinfo  {journal} {Phys. Rev. Lett.}\ }\textbf {\bibinfo {volume} {122}},\
  \bibinfo {pages} {020502} (\bibinfo {year} {2019})}\BibitemShut {NoStop}%
\bibitem [{\citenamefont {Bausch}(2022)}]{Bausch2022}%
  \BibitemOpen
  \bibfield  {author} {\bibinfo {author} {\bibfnamefont {J.}~\bibnamefont
  {Bausch}},\ }\href {https://doi.org/10.22331/q-2022-08-04-773} {\bibfield
  {journal} {\bibinfo  {journal} {{Quantum}}\ }\textbf {\bibinfo {volume}
  {6}},\ \bibinfo {pages} {773} (\bibinfo {year} {2022})}\BibitemShut {NoStop}%
\bibitem [{\citenamefont {Lubasch}\ \emph {et~al.}(2020)\citenamefont
  {Lubasch}, \citenamefont {Joo}, \citenamefont {Moinier}, \citenamefont
  {Kiffner},\ and\ \citenamefont {Jaksch}}]{PhysRevA.101.010301}%
  \BibitemOpen
  \bibfield  {author} {\bibinfo {author} {\bibfnamefont {M.}~\bibnamefont
  {Lubasch}}, \bibinfo {author} {\bibfnamefont {J.}~\bibnamefont {Joo}},
  \bibinfo {author} {\bibfnamefont {P.}~\bibnamefont {Moinier}}, \bibinfo
  {author} {\bibfnamefont {M.}~\bibnamefont {Kiffner}},\ and\ \bibinfo {author}
  {\bibfnamefont {D.}~\bibnamefont {Jaksch}},\ }\href
  {https://doi.org/10.1103/PhysRevA.101.010301} {\bibfield  {journal} {\bibinfo
   {journal} {Phys. Rev. A}\ }\textbf {\bibinfo {volume} {101}},\ \bibinfo
  {pages} {010301} (\bibinfo {year} {2020})}\BibitemShut {NoStop}%
\bibitem [{\citenamefont {Plekhanov}\ \emph {et~al.}(2022)\citenamefont
  {Plekhanov}, \citenamefont {Rosenkranz}, \citenamefont {Fiorentini},\ and\
  \citenamefont {Lubasch}}]{Plekhanov2022variationalquantum}%
  \BibitemOpen
  \bibfield  {author} {\bibinfo {author} {\bibfnamefont {K.}~\bibnamefont
  {Plekhanov}}, \bibinfo {author} {\bibfnamefont {M.}~\bibnamefont
  {Rosenkranz}}, \bibinfo {author} {\bibfnamefont {M.}~\bibnamefont
  {Fiorentini}},\ and\ \bibinfo {author} {\bibfnamefont {M.}~\bibnamefont
  {Lubasch}},\ }\href {https://doi.org/10.22331/q-2022-03-17-670} {\bibfield
  {journal} {\bibinfo  {journal} {{Quantum}}\ }\textbf {\bibinfo {volume}
  {6}},\ \bibinfo {pages} {670} (\bibinfo {year} {2022})}\BibitemShut {NoStop}%
\bibitem [{\citenamefont {Akhalwaya}\ \emph {et~al.}(2023)\citenamefont
  {Akhalwaya}, \citenamefont {Connolly}, \citenamefont {Guichard},
  \citenamefont {Herbert}, \citenamefont {Kargi}, \citenamefont {Krajenbrink},
  \citenamefont {Lubasch}, \citenamefont {Keever}, \citenamefont {Sorci},
  \citenamefont {Spranger},\ and\ \citenamefont
  {Williams}}]{akhalwaya2023modularenginequantummonte}%
  \BibitemOpen
  \bibfield  {author} {\bibinfo {author} {\bibfnamefont {I.~Y.}\ \bibnamefont
  {Akhalwaya}}, \bibinfo {author} {\bibfnamefont {A.}~\bibnamefont {Connolly}},
  \bibinfo {author} {\bibfnamefont {R.}~\bibnamefont {Guichard}}, \bibinfo
  {author} {\bibfnamefont {S.}~\bibnamefont {Herbert}}, \bibinfo {author}
  {\bibfnamefont {C.}~\bibnamefont {Kargi}}, \bibinfo {author} {\bibfnamefont
  {A.}~\bibnamefont {Krajenbrink}}, \bibinfo {author} {\bibfnamefont
  {M.}~\bibnamefont {Lubasch}}, \bibinfo {author} {\bibfnamefont {C.~M.}\
  \bibnamefont {Keever}}, \bibinfo {author} {\bibfnamefont {J.}~\bibnamefont
  {Sorci}}, \bibinfo {author} {\bibfnamefont {M.}~\bibnamefont {Spranger}},\
  and\ \bibinfo {author} {\bibfnamefont {I.}~\bibnamefont {Williams}},\ }\href
  {https://arxiv.org/abs/2308.06081} {\bibinfo {title} {A modular engine for
  quantum monte carlo integration}} (\bibinfo {year} {2023}),\ \Eprint
  {https://arxiv.org/abs/2308.06081} {arXiv:2308.06081 [quant-ph]} \BibitemShut
  {NoStop}%
\bibitem [{\citenamefont {Iaconis}\ \emph {et~al.}(2024)\citenamefont
  {Iaconis}, \citenamefont {Johri},\ and\ \citenamefont {Zhu}}]{Iaconis2024}%
  \BibitemOpen
  \bibfield  {author} {\bibinfo {author} {\bibfnamefont {J.}~\bibnamefont
  {Iaconis}}, \bibinfo {author} {\bibfnamefont {S.}~\bibnamefont {Johri}},\
  and\ \bibinfo {author} {\bibfnamefont {E.~Y.}\ \bibnamefont {Zhu}},\ }\href
  {https://doi.org/10.1038/s41534-024-00805-0} {\bibfield  {journal} {\bibinfo
  {journal} {npj Quantum Inf.}\ }\textbf {\bibinfo {volume} {10}},\ \bibinfo
  {pages} {15} (\bibinfo {year} {2024})}\BibitemShut {NoStop}%
\bibitem [{\citenamefont {Sano}\ and\ \citenamefont
  {Hamamura}(2024)}]{sano2024}%
  \BibitemOpen
  \bibfield  {author} {\bibinfo {author} {\bibfnamefont {Y.}~\bibnamefont
  {Sano}}\ and\ \bibinfo {author} {\bibfnamefont {I.}~\bibnamefont
  {Hamamura}},\ }\href {https://arxiv.org/abs/2403.16729} {\bibinfo {title}
  {Quantum state preparation for probability distributions with mirror symmetry
  using matrix product states}} (\bibinfo {year} {2024}),\ \Eprint
  {https://arxiv.org/abs/2403.16729} {arXiv:2403.16729 [quant-ph]} \BibitemShut
  {NoStop}%
\bibitem [{\citenamefont {Sun}\ \emph {et~al.}(2023)\citenamefont {Sun},
  \citenamefont {Tian}, \citenamefont {Yang}, \citenamefont {Yuan},\ and\
  \citenamefont {Zhang}}]{Sun2023}%
  \BibitemOpen
  \bibfield  {author} {\bibinfo {author} {\bibfnamefont {X.}~\bibnamefont
  {Sun}}, \bibinfo {author} {\bibfnamefont {G.}~\bibnamefont {Tian}}, \bibinfo
  {author} {\bibfnamefont {S.}~\bibnamefont {Yang}}, \bibinfo {author}
  {\bibfnamefont {P.}~\bibnamefont {Yuan}},\ and\ \bibinfo {author}
  {\bibfnamefont {S.}~\bibnamefont {Zhang}},\ }\href
  {https://doi.org/10.1109/TCAD.2023.3244885} {\bibfield  {journal} {\bibinfo
  {journal} {IEEE Transactions on Computer-Aided Design of Integrated Circuits
  and Systems}\ }\textbf {\bibinfo {volume} {42}},\ \bibinfo {pages} {3301}
  (\bibinfo {year} {2023})}\BibitemShut {NoStop}%
\bibitem [{\citenamefont {Zhang}\ \emph {et~al.}(2022)\citenamefont {Zhang},
  \citenamefont {Li},\ and\ \citenamefont {Yuan}}]{Zhang2022}%
  \BibitemOpen
  \bibfield  {author} {\bibinfo {author} {\bibfnamefont {X.-M.}\ \bibnamefont
  {Zhang}}, \bibinfo {author} {\bibfnamefont {T.}~\bibnamefont {Li}},\ and\
  \bibinfo {author} {\bibfnamefont {X.}~\bibnamefont {Yuan}},\ }\href
  {https://doi.org/10.1103/PhysRevLett.129.230504} {\bibfield  {journal}
  {\bibinfo  {journal} {Phys. Rev. Lett.}\ }\textbf {\bibinfo {volume} {129}},\
  \bibinfo {pages} {230504} (\bibinfo {year} {2022})}\BibitemShut {NoStop}%
\bibitem [{\citenamefont {Rattew}\ and\ \citenamefont
  {Koczor}(2022)}]{Rattew2022}%
  \BibitemOpen
  \bibfield  {author} {\bibinfo {author} {\bibfnamefont {A.~G.}\ \bibnamefont
  {Rattew}}\ and\ \bibinfo {author} {\bibfnamefont {B.}~\bibnamefont
  {Koczor}},\ }\href {https://arxiv.org/abs/2205.00519} {\bibinfo {title}
  {Preparing arbitrary continuous functions in quantum registers with
  logarithmic complexity}} (\bibinfo {year} {2022}),\ \Eprint
  {https://arxiv.org/abs/2205.00519} {arXiv:2205.00519 [quant-ph]} \BibitemShut
  {NoStop}%
\bibitem [{\citenamefont {McArdle}\ \emph {et~al.}(2022)\citenamefont
  {McArdle}, \citenamefont {Gilyén},\ and\ \citenamefont
  {Berta}}]{McArdle2022}%
  \BibitemOpen
  \bibfield  {author} {\bibinfo {author} {\bibfnamefont {S.}~\bibnamefont
  {McArdle}}, \bibinfo {author} {\bibfnamefont {A.}~\bibnamefont {Gilyén}},\
  and\ \bibinfo {author} {\bibfnamefont {M.}~\bibnamefont {Berta}},\ }\href
  {https://arxiv.org/abs/2210.14892} {\bibinfo {title} {Quantum state
  preparation without coherent arithmetic}} (\bibinfo {year} {2022}),\ \Eprint
  {https://arxiv.org/abs/2210.14892} {arXiv:2210.14892 [quant-ph]} \BibitemShut
  {NoStop}%
\bibitem [{\citenamefont {Cormen}\ \emph {et~al.}(2022)\citenamefont {Cormen},
  \citenamefont {Leiserson}, \citenamefont {Rivest},\ and\ \citenamefont
  {Stein}}]{Cormen2009}%
  \BibitemOpen
  \bibfield  {author} {\bibinfo {author} {\bibfnamefont {T.~H.}\ \bibnamefont
  {Cormen}}, \bibinfo {author} {\bibfnamefont {C.~E.}\ \bibnamefont
  {Leiserson}}, \bibinfo {author} {\bibfnamefont {R.~L.}\ \bibnamefont
  {Rivest}},\ and\ \bibinfo {author} {\bibfnamefont {C.}~\bibnamefont
  {Stein}},\ }\href@noop {} {\emph {\bibinfo {title} {Introduction to
  algorithms}}},\ \bibinfo {edition} {4th}\ ed.\ (\bibinfo  {publisher} {The
  MIT Press},\ \bibinfo {year} {2022})\ pp.\ \bibinfo {pages}
  {76--125}\BibitemShut {NoStop}%
\bibitem [{\citenamefont {M\"{o}tt\"{o}nen}\ \emph {et~al.}(2005)\citenamefont
  {M\"{o}tt\"{o}nen}, \citenamefont {Vartiainen}, \citenamefont {Bergholm},\
  and\ \citenamefont {Salomaa}}]{Mottonen2005}%
  \BibitemOpen
  \bibfield  {author} {\bibinfo {author} {\bibfnamefont {M.}~\bibnamefont
  {M\"{o}tt\"{o}nen}}, \bibinfo {author} {\bibfnamefont {J.~J.}\ \bibnamefont
  {Vartiainen}}, \bibinfo {author} {\bibfnamefont {V.}~\bibnamefont
  {Bergholm}},\ and\ \bibinfo {author} {\bibfnamefont {M.~M.}\ \bibnamefont
  {Salomaa}},\ }\href {https://dl.acm.org/doi/10.5555/2011670.2011675}
  {\bibfield  {journal} {\bibinfo  {journal} {Quantum Info. Comput.}\ }\textbf
  {\bibinfo {volume} {5}},\ \bibinfo {pages} {467–473} (\bibinfo {year}
  {2005})}\BibitemShut {NoStop}%
\bibitem [{\citenamefont {Araujo}\ \emph {et~al.}(2021)\citenamefont {Araujo},
  \citenamefont {Park}, \citenamefont {Petruccione},\ and\ \citenamefont
  {da~Silva}}]{Araujo2021}%
  \BibitemOpen
  \bibfield  {author} {\bibinfo {author} {\bibfnamefont {I.~F.}\ \bibnamefont
  {Araujo}}, \bibinfo {author} {\bibfnamefont {D.~K.}\ \bibnamefont {Park}},
  \bibinfo {author} {\bibfnamefont {F.}~\bibnamefont {Petruccione}},\ and\
  \bibinfo {author} {\bibfnamefont {A.~J.}\ \bibnamefont {da~Silva}},\ }\href
  {https://doi.org/10.1038/s41598-021-85474-1} {\bibfield  {journal} {\bibinfo
  {journal} {Sci. Rep.}\ }\textbf {\bibinfo {volume} {11}},\ \bibinfo {pages}
  {6329} (\bibinfo {year} {2021})}\BibitemShut {NoStop}%
\bibitem [{\citenamefont {Park}\ \emph {et~al.}(2019)\citenamefont {Park},
  \citenamefont {Sinayskiy}, \citenamefont {Fingerhuth}, \citenamefont
  {Petruccione},\ and\ \citenamefont {Rhee}}]{Park2019b}%
  \BibitemOpen
  \bibfield  {author} {\bibinfo {author} {\bibfnamefont {D.~K.}\ \bibnamefont
  {Park}}, \bibinfo {author} {\bibfnamefont {I.}~\bibnamefont {Sinayskiy}},
  \bibinfo {author} {\bibfnamefont {M.}~\bibnamefont {Fingerhuth}}, \bibinfo
  {author} {\bibfnamefont {F.}~\bibnamefont {Petruccione}},\ and\ \bibinfo
  {author} {\bibfnamefont {J.-K.~K.}\ \bibnamefont {Rhee}},\ }\href
  {https://doi.org/10.1088/1367-2630/ab35fb} {\bibfield  {journal} {\bibinfo
  {journal} {New J. of Phys.}\ }\textbf {\bibinfo {volume} {21}},\ \bibinfo
  {pages} {083024} (\bibinfo {year} {2019})}\BibitemShut {NoStop}%
\bibitem [{\citenamefont {Kitaev}\ and\ \citenamefont
  {Webb}(2009)}]{kitaev2009wavefunction}%
  \BibitemOpen
  \bibfield  {author} {\bibinfo {author} {\bibfnamefont {A.}~\bibnamefont
  {Kitaev}}\ and\ \bibinfo {author} {\bibfnamefont {W.~A.}\ \bibnamefont
  {Webb}},\ }\href@noop {} {\bibinfo {title} {Wavefunction preparation and
  resampling using a quantum computer}} (\bibinfo {year} {2009}),\ \Eprint
  {https://arxiv.org/abs/0801.0342} {arXiv:0801.0342 [quant-ph]} \BibitemShut
  {NoStop}%
\bibitem [{Note1()}]{Note1}%
  \BibitemOpen
  \bibinfo {note} {If $\zeta $ were varied continuously from 0 to $1/2^{n-1}$
  then ergodic coverage of the entire PDF over the breadth of the sample window
  would be achieved.}\BibitemShut {Stop}%
\bibitem [{\citenamefont {da~Silva/Centro~de
  Inform\'atica~UFPE}(2020)}]{daSilvaCode2020}%
  \BibitemOpen
  \bibfield  {author} {\bibinfo {author} {\bibfnamefont {A.~J.}\ \bibnamefont
  {da~Silva/Centro~de Inform\'atica~UFPE}},\ }\href@noop {} {\bibinfo {title}
  {Github---adjs/dcsp}},\ \bibinfo {howpublished}
  {\url{https://github.com/adjs/dcsp}} (\bibinfo {year} {July 2020}),\ \bibinfo
  {note} {accessed: November 2024}\BibitemShut {NoStop}%
\end{thebibliography}%

\end{document}